\documentstyle[sprocl,epsf,art12]{article}
\def\Journal#1#2#3#4{{#1} {\bf #2}, #3 (#4)}

\def\PGNP{\em J.Phys. G.Nucl. Phys.}
\def\NPB{{\em Nucl. Phys.} B}
\def\NPA{{\em Nucl. Phys.} A}

\def\ZPC{{\em Z. Phys.} C}
\begin{document}
\vspace*{1.cm}
\begin{center}
\begin{Large}\begin{bf}
Double Phase Transition Model and the problem of entropy and baryon number
conservation.
\end{bf}\end{Large}
\end{center}
~\\~\\
\hspace*{1.54cm}{\large {\bf O.D.Chernavskaya}}
~\\
\hspace*{1.54cm}{\bf  P.N.Lebedev Physical Institute, Moscow}\\
\hspace*{1.54cm} 117333, Leninski prospect, 53 \\
\hspace*{1.54cm}e-mail:  chernav@lpi.ac.ru\\
~\\~\\
\subsection*{Abstract}
\begin{small}
\hspace*{.9cm}We continue to develop further the bag-type
Double Phase Transition Model (DPTM)
for transformation of Quark-Gluon Plasma ($QGP$) to normal
hadronic matter ($H$-phase). The model is based on the assumed existence
of an {\it intermediate Q phase} composed by massive {\it constituent quarks}
and pions (as Goldstone bosons) \cite{divonne}$^,$ \cite{itep}.

In the present paper the problem of entropy $\cal S$ and baryon number
$\cal N_ B$ conservation in phase transitions from deconfined phases
($QGP$ and $Q$) to hadronic matter $H$ is considered.
It is shown that standard construction of both first order
phase transitions, $H \leftrightarrow Q$ as well as $Q \leftrightarrow QGP$
implies a discontinuous structure of entropy per baryon $\cal S/N_B$
when crossing phase boundary;
this results in impossibility of equilibrium transition from $QGP$ to
hadron gas.

We follow the way suggested recently \cite{satz}$^,$ \cite{bilef} by
H.Satz et al. for the same problem concerning direct transition
$H\leftrightarrow QGP$.
They proposed a modification of bag pressure
parameter $B_{QGP}$ by making it dependent on system temperature $T$ and
baryon chemical potential $\mu $; this modification has been
demonstrated to be sufficient to provide $\cal S/N_B$ conservation.

Here we show that within DPTM such a modification turns out
to be necessary and sufficient for bag pressure $B_Q$ in the
$Q$ phase {\it only}.
The DPTM modified in such a way is shown to satisfy equilibrium Gibbs
criteria for phase transitions. Location of phase boundaries in $\mu -T$
plane  has been demonstrated to be changed but slightly;  the modification
tells mainly on baryon number density within $Q$ phase.
Two alternative descriptions of nucleon-nucleon
interaction -the Hard Core Model and the Mean Field Approximation -
have been tested; the results for both cases appeared to be similar.
All the results are shown to be stable against rather broad variations of
model parameters.
\end{small}
 \newpage
\section{Introduction}
\hspace*{.9cm} According to the fundamental QCD predictions strongly
interacting matter has to exist in (at least) two different phases \cite{qm96}
:\\
- Quark-Gluon Plasma ($QGP$), i.e. gas of  deconfined and massless quarks $q$,
antiquarks $\bar q$ and gluons $g$, - at super high temperature $T$ and
energy density  $\epsilon $, and \\
-Hadron Gas , or $H$-phase at low $T$ and/or $\epsilon $.

QCD lattice calculations \cite{lattice} confirm this concept indicating to
the first order phase transition at $T \approx$ 200 MeV.
Well above this temperature matter behaviour is easily
described by QCD perturbation theory. Close to the critical
conditions perturbation theory fails.
However critical behaviour of strongly interacting
matter is very important to analyze the phase transition conditions
and the $QGP$ signals.
To study this problem one has to use some phenomenological models.
Thermodynamic approach with two-phase matter Equations of State (EOS) has been
 widely used recently \cite{thermo} and gave a lot of interesting results and
experimental predictions.  However there still exist some ambiguous points
connected with phase transition description. One of them
is discussed here.

We consider, following the common way \cite{rev},
bag type EOS for $QGP$ and usual nonrelativistic EOS for $H$ phase
(taking into account hadron interaction).
Such type models \cite{thermo} are known to result in first order phase
transition $H \leftrightarrow QGP$ at some critical temperature $T_c$,
{\it if none intermediate phase is taken into account} (to be called later on
Single Phase Transition Model, or SPTM).
However, in distinction to the great majority of researchers considering
{\it only direct} phase transition $H \leftrightarrow QGP$, and thus assuming
that, both, quark deconfinement and chiral symmetry
restoration have to occur simultaneously, we have worked out
\cite{divonne}$^,$ \cite{itep}
the  Double Phase Transition Model (DPTM) based on the assumption of
an intermediate phase $Q$ formed by {\it deconfined constituent}
quarks and pions (as necessary Goldstone bosons). In the form
proposed in \cite{divonne}$^,$ \cite{itep} it was actually demonstrated that
such a $Q$ phase could exist.

The idea of possibility of two phase transitions goes back to E. Shuryak
\cite{shur} who put forward various arguments to point out that
temperatures of
quark deconfinement $T_d$ and chiral symmetry restoration $T_{ch}$
{\it may not coincide}, and $T_d$ should be {\it less} than $T_{ch}$.
Later this idea was supported by other works within QCD
\cite{elf}$^,$ \cite{qcd-2pt}.
It leads to the conclusion that there may exist some
temperature interval $T_d<T<T_{ch}$ where quarks are liberated off
individual hadrons but still possess non-zero mass. Such objects are well
known from Additive Quark Model (AQM) and called {\it constituent
quarks}, or {\it valons}. They are necessary entities for satisfactory
description of moderate energy hadron phenomena \cite{AQM}. Thus an
intermediate phase of deconfined massive constituent quarks, $Q$-phase,
may exist.

First attempt to investigate this problem within the thermodynamic models
with bag type EOS belongs to the Bielefeld group \cite{2-4-1}$^,$ \cite{2-4-2};
then this problem has been investigated in Refs.\cite{2-4-3}$^-$\cite{2-4-5}.
A possibility for existence of
the intermediate phase $Q$ formed by deconfined constituent quarks and pions
 was in fact demonstrated. However the choice of the
key model parameters based on the lattice calculation data for baryonless
matter \cite{lattice}
resulted in the negative conclusion that $H\leftrightarrow QGP$ transition
should proceed almost always
directly, without any intermediate state, since
in temperature $T$ - chemical potential $\mu$ plane  $Q$ phase
occupies only  tiny petals and can hardly play any essential role in reality.

In our earlier works \cite{divonne}$^,$ \cite{itep} the problem has been reconsidered
following the same ideology \cite{2-4-1}$^-$ \cite{2-4-5}  but with different
physical approach to the choice of the bag parameters, since lattice approach
(not securing pion description) does not seem to be a safe basis in the
case of $H \leftrightarrow Q$ transformation where pions play the
decisive role.

This resulted in conclusion that $Q$ phase seemingly exists almost always
(at least, for $\mu \le 1$). In $\mu -T$ plane it occupies a corridor between
$H$ and $QGP$ phases having the width $\Delta T=T_{ch}-T_d \approx $ 50 MeV.
The value of $T_d$ was found to be equal to some 150 MeV and its physical
meaning (the highest temperature allowing for existence of hadrons, above
$T_d$ they have to decay into constituent quarks) enables to identify it
with the Hagedorn temperature \cite{haged}. Both phase transitions are of the
first order.

The qualitative stability of these results for varying values of parameters
(within reasonable limits)  has been demonstrated. More details are given in
the  section 2.

The present paper is devoted to the problem of entropy and baryon number
conservation when crossing phase transition boundaries. This problem
appeared already in the common model with single (first order)
phase transition, SPTM.
It has been shown \cite{satz}$^,$ \cite{bilef}
within SPTM  that as a result of the
transition $H \leftrightarrow QGP$
the entropy per baryon ${\cal S/N_B}$
is discontinuous  across the phase transition boundary:
\begin{center}
$({\cal S/N_B})_{QGP} \quad > \quad ({\cal S/N_B})_{H}$.
\end{center}
This means that transition $H \leftrightarrow QGP$ {\it is irreversible}:
it could not satisfy thermal and chemical equilibrium conditions, and,
at the same time, fulfill baryon number and entropy
conservation at the phase boundary. In particular, adiabatic transition
$QGP \rightarrow H$ is impossible as it should be accompanied
by the entropy {\it decrease}
(since conservation of the baryon number is secured).

Thus one has to choose between two possibilities:\\
-either the transition from $QGP$ into $H$-phase
{\it can not proceed as an adiabatic one} under any conditions, \\
- or EOS used {\it are not fully correct}.

The first possibility has been discussed recently \cite{entro} in connection
with large value of specific entropy
$({\cal S/N_B})_{H}$ detected in experiments on
high energy heavy ion collisions. It was used as an argument for
assumption that $QGP$ has been actually observed.

The later possibility seems rather realistic due to  uncertainties
presenting in phenomenological EOS. In particular, bag-type
EOS used for $QGP$ includes the key model
parameter $B_{\small QGP}$ (representing nonperturbative
interactions of quarks and gluons with physical vacuum).
In various works on SPTM its value was chosen rather arbitrary:
$B_{\small QGP}$ = 0.2 $\div $ 0.5 GeV/fm$^3$.
Usually it is treated as constant, but there are no reasons for $B$
not to depend on $T$ and/or $\mu $.

Accordingly in \cite{bilef} there was suggested a certain modification of $B_{\small QGP}$
making it $\mu$ and $T$ dependent,
$B(\mu, T)$, instead of commonly used $B_{QGP}=const$. This enabled to
restore the continuity of ${\cal S/N_B}$ and thus to  solve the problem
immediately. This modification of EOS was shown to change phase diagram of
the system {\it not considerably}, while the transition
$QGP \leftrightarrow H$ becomes reversible.


In this paper we study the problem of specific entropy conservation
within DPTM.  There appears the same discontinuity
of ${\cal S/N_B}$ when crossing {\it both transition boundaries}, thus
{\it both} phase transitions appear to be {\it irreversible}.
It is shown that the reversibility of both phase transitions can be
restored by {\it modifying the $Q$ phase EOS alone},
without changing $B_{\small QGP}$.

The paper is organized as follows. In section 2 we remind basic features
of the DPTM. Specific entropy discontinuity and the method of its correction
within SPTM is discussed in section 3; in section 4 the same problem is
discussed within DPTM; results of numerical calculations are presented in
subsection. Section 5 represents summary and discussion.
Some details of hydrodynamical description of equilibrium system evolution
are given in Appendix.
\section{Basic features and main results of DPTM}
\hspace*{.9cm} We use, following common way \cite{2-4-1}$^-$ \cite{2-4-5},
bag-type model EOS for $QGP$ and $Q$ phases:
($p, g_{i}$, $m_{i}$  are  pressure, degeneracy
factors, masses  of $i$-th type
particles respectively for each $j$-th phase;
$j$ means hadronic $H$,valonic $Q$, and $QGP$ phases respectively).
\begin{equation}
p_{QGP}(T,\mu,V) = \frac {\pi^2}{90}(g_g+g_q\frac{7}{4})T^4~+~ T^2\mu^2_q~+~
\frac {1}{2\pi^2}\mu^4_q~-~B_{QGP},
\end{equation}
\begin{eqnarray}
\hspace*{-.75cm}&&p_Q (T,\mu,V) = \Biggl\{
\frac{g_\pi}{6\pi^2} \int \frac{k^4 dk}{\sqrt{k^2+m^2_\pi}}~
\frac{1}{exp\left(\frac{\sqrt{k^2+m^2_\pi }}{T}\right) - 1}
\quad +  \nonumber\\[-2mm]
\hspace*{-.75cm}&&\\[-2mm]
\hspace*{-.75cm}&&\sum_{i=u,d,s} \frac{g_i}{6\pi^2} \int \frac{k^4 dk}{\sqrt{k^2+m^2_i}} \left[\
\frac{1}{\exp\left( \frac{\sqrt{k^2+m^2_i}-\mu^i_Q}{T}\right)+1} +
\frac{1}{\exp\left( \frac{\sqrt{k^2+m^2_i}+\mu^i_Q}{T}\right) +1} \right]\
\Biggr\}-B_Q
\nonumber
\label{Q}
\end{eqnarray}
\par Massless gluons ($g$) and quarks ($q$) participate in $QGP$;
$Q$ phase contains constituent quarks ($u,d,s$ with
$m_{u}\simeq m_d\simeq $320
MeV and  $m_s\simeq$ 512 MeV), and pions;
%
$\mu^{(i)}_{Q}$ being chemical potential of constituent quarks,
equal to that of corresponding current quarks ($\mu_q=\mu_u=\mu_d$, and
$\mu_s$ is taken to be zero).

The terms  $B_j$  in EOS of $Q$ and $QGP$ reflect effective interactions
with vacuum. $B_{Q}, B_{QGP}$ are free bag parameters chosen according to their
physical meaning: $B_{QGP}$ is QCD vacuum energy density known \cite{bvac} to
be $\approx$ 0.5 $\div$ 1 GeV/fm$^3$,  and $B_Q$ is estimated from
low energy phenomena \cite{AQM} as 50 $\div$ 100 MeV/fm$^3$
(close to the B value of MIT-bag model).
Note that this choice of $B$ values (instead of that chosen in
\cite{2-4-1}$^-$ \cite{2-4-5}) is crucial for appearance of the
intermediate $Q$ phase.

For the hadronic phase $H$ :
\begin{eqnarray}
\hspace*{-.5cm}&& p_H(T,\mu,V) = \frac{g_\pi}{6\pi^2} \int \frac{k^4 dk}{\sqrt{k^2+m^2_\pi}}~
\frac{1}{exp{\left( \frac{\sqrt{k^2+m^2_\pi}}{T} \right)} - 1}
\quad + \nonumber\\[-2mm]
\hspace*{-.5cm}&&\\[-2mm]
\hspace*{-.5cm}&&\sum_i \frac{g_i}{6\pi^2} \int \frac{k^4 dk}{\sqrt
{k^2+m^2_i}} \left[\frac{1}{\exp\left( \frac{\sqrt{k^2+m^2_i}-\mu_i}{T}
\right)+1} + \frac{1}{\exp\left( \frac{\sqrt{k^2+m^2_i}+\mu_i}{T}\right) +1}
\right]+ \phi (U(\nu))
\nonumber
\label{H}
\end{eqnarray}
~\\
The first term in (3) represents pion contribution;
the summation (in the second term) is over all $i-th$ type stable hadrons
dominating in the $H$ phase ($\pi$, $N$, $\Lambda$ and $K$ were taken into account);
$m_i$ , $\mu_i$ and $g_i$ are the corresponding masses, chemical potentials
and degeneracy factors. The last term
stands for account of nucleon-nucleon interactions in the form of
Mean Field Approximation (MFA) \cite{mfa}; we have
used also Hard Core Model (HCM) for nucleon interaction description
\cite{2-4-1}$^,$ \cite{2-4-2}
(where the form of interaction account in the $H$ phase is more
complicated).

In thermodynamic equilibrium, according to Gibbs principle,
 the actually realized phase is that with the largest pressure
at given $\mu$ and $T$.
Then at given value of nucleon chemical potential, $\mu$
there are {\it 3 possible transition temperatures}:

{\it  deconfinement transition} ($H\leftrightarrow Q$) at $T_d$:  \qquad
$p_H(T_d, \mu)=p_Q(T_d, \mu/3)$

{\it  direct transition}  ($H \leftrightarrow QGP$) at $T_c$: \quad \qquad
 $p_H(T_c, \mu)=p_{QGP}(T_c, \mu/3)$

{\it  chiral transition} ($Q\leftrightarrow QGP$) at $T_{ch}$: \quad \qquad
$p_Q(T_{ch}, \mu/3)=p_{QGP}(T_{ch}, \mu/3)$

The case of {\it coincidence} of all three transitions
corresponds to the
{\it triple point} at $T^\#$: \qquad \qquad
$p_H(T^\#, \mu)=p_{Q}(T^\#, \mu/3)=p_{QGP}(T^\#, \mu/3)$

$Q$ phase actually exists if for rising temperature the deconfinement of
valons occurs first,
prior to direct formation of the $QGP$ phase, i.e. $T_d < T_c$; and,
in the opposite direction (for decreasing temperature), if chiral transition
from $QGP$ to $Q$ phase
occurs prior to formation of $H$ phase, i.e.  $T_{ch} > T_c$. Thus
general condition for $Q$ phase  existence is:
 \begin{equation}
T_d(\mu)<T_{c}(\mu)<T_{ch}(\mu).
\end{equation}
In this case the {\it direct transition does not occur}.
Otherwise DPTM reduces to a model with single phase
transition and its results coincide with  that of SPTM.
This very case have been met in the papers \cite{2-4-1}$^-$ \cite{2-4-5}
due to specific choice of the $B$ parameters.

The choice of model parameters based on its physical meaning
resulted in the quite opposite conclusions, namely
\cite{divonne}$^,$ \cite{itep}:\\
\noindent {\bf i)} $H\leftrightarrow QGP$ transition proceeds almost
exclusively via the $Q$ phase, $H\leftrightarrow Q\leftrightarrow QGP$.\\
\noindent {\bf ii)} Deconfinement of valons
$H\rightarrow Q$ should occur at rather
low energy density of nuclear matter $\approx $ 0.3$\div$0.4 GeV/fm$^3$
(only three times larger than energy density in a normal nucleus,
as it was roughly estimated in \cite{elf}).\\
\noindent {\bf iii)} Temperature interval for $Q$ phase, $\Delta
T=T_{ch}-T_{d}$, amounts to $\sim$ 50 MeV (see Fig. 1).
For baryonless matter, typically, $T_d \approx$ 140 MeV and
$T_{ch} \approx$ 200 MeV.
Thus $T_d$ coincides with the well known Hagedorn temperature
(as it should be since at $T_d$ hadrons cease to exist and decay into
constituent quarks),
which had been treated earlier as some approximation to direct
transition $H\leftrightarrow QGP$ temperature \cite{haged} and now is
shown to have actually independent physical meaning.

These results turned out to be qualitatively stable against extended
variations of model parameters and nucleon interaction description.

\section{ Restoration of specific entropy continuity in SPTM.}
\hspace*{.9cm} As it has been said above, within SPTM the specific entropy value turns
out to be discontinuous when crossing the direct transition boundary (
the ratio of specific entropy values above and below the direct transition
are presented in Fig. 2a).  However, it was shown \cite{bilef} that the value $({\cal
S/N_B})_{QGP}$ can be corrected by the modification:  $B_{QGP} \rightarrow
B_{QGP}(\mu,T)$.  Indeed, according to general thermodynamic relations
entropy density $s$ and baryon number density $n$ in $QGP$ are defined from:
\begin{equation}
n_{QGP}(\mu,T) \equiv \frac{\partial p_{QGP}(\mu,T)}{\partial \mu} =
n^0_{QGP}(\mu,T) - \frac{\partial B_{QGP}(\mu,T)}{\partial \mu},
\end{equation}
\begin{equation}
s_{QGP} (\mu,T) \equiv  \frac{\partial p_{QGP}(\mu,T,)}{\partial T} =
s^0_{QGP}(\mu,T) - \frac {\partial B_{QGP}(\mu,T)}{\partial T},
\end{equation}
where zero superscripts indicate corresponding values calculated for
constant $B$.
Other thermodynamic functions in $QGP$ ($\epsilon $, $p$, enthalpy $w$)
remain undependent on $B$'s derivatives.

Thus to restore conservation of specific entropy
it is possible to determine $B_{QGP}(\mu,T)$ from the differential equation:
\begin{equation}
\frac{s^0_{QGP}(\mu,T) - \frac {\partial B_{QGP}(\mu,T)}{\partial T}}
{n^0_{QGP}(\mu,T) - \frac{\partial B_{QGP}(\mu,T)}{\partial \mu}} =
\frac{s_H(\mu,T)}{n_H(\mu,T)}.
\end{equation}
Since ${\cal S/N_B}$ value is not defined at the points $\mu=0$ and $T=0$
it seems natural to fix an integrating constant:
\begin{center}
$B_{QGP}(0,T)=B_{QGP}(\mu,0) = B^0_{QGP}$.
\end{center}

Let us stress that the
function $B_{QGP}(\mu,T)$ satisfying eq. (7) provides equal behaviour of
specific entropy functions in both phases {\it everywhere in  $\mu -T$ plane},
thus, in particular, conservation of the specific entropy at the phase
transition boundary.

This equation has been solved in Ref.\cite{bilef}; it was shown that
the  obtained $B_{QGP}(\mu,T)$  does not change phase diagram considerably.

However, there still remain several questions concerning this
procedure. In particular,
$B_{QGP}(\mu,T)$, as a solution of the eq. (7), is defined in the open
region in $\mu -T$ plane, i.e. for any temperature $T>T_c(\mu)$. Since it
depends on the value of specific entropy in $H$ phase, it means that
EOS of $QGP$ (in particular, the bag constant) should store
information on the $H$-phase EOS everywhere including far high-$T$ limit.
This looks suspicious since $B_{QGP}$ represents pressure of the physical
vacuum and has nothing in common with a particular hadronic system.

This question becomes serious at relatively large $\mu \ge $ 0.9 GeV/fm$^3$
where hadron interactions
become decisive in the $H$ phase \cite{itep}.
One has to use again phenomenological models for description of those
interactions (eg., HCM and MFA) which give differing results for the
${\cal (S/N_B)}_{H}$ value (see Fig. 2a).
Correspondingly, $B_{QGP}(\mu,T)$ modified according to the
procedure described above {\it should
 depend on description of nucleon interactions in the $H$ phase}.

Besides, there appears some uncertainty concerning fulfillment of Gibbs
relation
\begin{equation}
\epsilon +p -\mu n = sT,
\end{equation}
which is to be satisfied in an equilibrium system. Note that this imposes
additional constraint on B's derivatives in the case of modified $B_{QGP}(\mu,
T)$. Indeed, combining (8) with (5),(6) one gets the condition:
\begin{center}
$\mu
\frac {\partial B(\mu, T)}{\partial \mu} = - T \frac {\partial
B(\mu, T)}{\partial T}$,
\end{center}
which, seemingly, is not satisfied within the
procedure used in \cite{bilef}.

Another problem concerns dynamical evolution of the system
in question.  The analysis presented refers to the stationary systems.  Being
related to modern experiments on high energy heavy ion collisions, it is to
be included in hydrodynamical model of the system evolution (for qualitative
analysis the simplest Bjorken scaling solution \cite{bjor} is used).  However
the condition (7) does not provide equilibrium character of first order phase
transition in dynamically evolving systems. Such transitions are to proceed
through the {\it mixed phase } state where not only entropy and baryon number
are to be conserved, as for all equilibrium processes, but {\it enthalpy}
of the system as well (the {\it latent heat} works on the system expansion).

To provide conservation of these three thermodynamic variables it is necessary
and sufficient (see Appendix A, eq.(27))
to fulfill the condition:
\begin{equation}
\frac {w_{QGP}(T_c, \mu/3)}{w_H(T_c,\mu)} =
\frac {s_{QGP}(T_c, \mu/3)}{s_H(T_c, \mu)} = \frac {n_{QGP}(T_c,
\mu/3)}{n_H(T_c,\mu)}.
\end{equation}
Thus, {\it both} $B_{QGP}(\mu ,T)$ derivatives are to be defined according to
(9),  {\it at least}, at the transition boundary $T_c(\mu)$.
Then fixing integrating constant in both ending points,
$\mu =0$ and $T=0$, as it was done in \cite{bilef}, makes the problem
over-defined.
\section{Entropy and baryon number conservation in DPTM}
\hspace{.9cm} The same problem arises also for transitions considered
within DPTM, i.e. for deconfinement and the chiral transition. The ratio
${\cal S} /{\cal N_B}$
turns out to be discontinuous when crossing {\it both} transition boundaries.
It is illustrated in Fig.2b, where "jumps" of  ${\cal S/N_B}$ ratio
at the corresponding transition boundaries are presented
as calculated, both, in the HCM and the MFA phenomenological models
describing nucleon interactions in the $H$ phase.

Following the same way as in \cite{bilef}, we try to reconsider EOS
for deconfined phase. However, within DPTM it seems natural and reasonable
to modify $Q$-{\it phase EOS alone} making the bag pressure parameter $B_Q$
$\mu $ and $T$ dependent, $B_Q(\mu ,T)$, with  $QGP$ bag parameter
$B_{QGP}$ remaining constant.

The reasons are as follows:\\
{\bf i).} $Q$-phase is the {\it intermediate} phase between $H$ and $QGP$,
thus $Q$ phase EOS can serve as a tool for compensation of defects of
other phases EOS (remaining the same as earlier)
which may occur invalid close to the transition boundaries.\\
{\bf ii).} $B_Q$ is chosen even more arbitrarily than $B_{QGP}$ : the last one is
to coincide with the QCD vacuum pressure estimated usually as 0.5 GeV/fm$^3$,
while the $B_Q$ is to be closed to the bag pressure within the MIT-bag model,
and thus varies within the interval: $B_Q \approx$ 50 $\div$ 100 MeV/fm$^3$.\\
{\bf iii).} The discontinuity of ${\cal S}/{\cal N_B}$ is {\it smaller} when crossing
the deconfinement and the chiral boundaries than that for direct transition,
(see Fig. 2),  thus it is easier to modify $Q$-phase EOS only.\\
{\bf iv).} $B_Q$ is defined for (and has a physical sense) {\it within closed region
of phase space} where the $Q$ phase can exist
(in $\mu -T$ plane: the region bounded by $T_d(\mu )$ and
$T_{ch}(\mu )$ curves and $\mu $=0 axis).
Thus there arises no problem with securing proper
$B_Q$ behaviour at high-$T$ limit,
as it appeared for $B_{QGP}(\mu ,T)$.\\
{\bf v).} modified $B_Q(\mu, T)$ would also depend on hadron interaction description,
and this dependence becomes considerable in high-$\mu$ limit. However
it is quite natural for $B_Q$ to be model dependent: EOS of the
{\it intermediate} phase has to vary in accordance with $H$-phase EOS
variations (in distinction to model dependence of $B_{QGP}(\mu ,T)$ in
SPTM which seems to be unnatural).\\

The modification in question, $B_Q(\mu,T)$,
results in change of entropy and baryon number density within $Q$ phase:
\begin{equation}
n_{Q}(\mu,T) \equiv \frac{\partial p_{Q}(\mu,T)}{\partial \mu} =
n^0_{Q}(\mu,T) - \frac{\partial B_{Q}(\mu,T)}{\partial \mu},
\end{equation}
\begin{equation}
s_{Q} (\mu,T) \equiv  \frac{\partial p_{Q}(\mu,T,)}{\partial T} =
s^0_{Q}(\mu,T) - \frac {\partial B_{Q}(\mu,T)}{\partial T},
\end{equation}
where zero superscripts indicate corresponding values calculated for
constant $B_Q$.
Other thermodynamic variables in $Q$-phase ($\epsilon_Q $, $p_Q$, $w_Q$)
remain unchanged, i.e. do not depend on $B_Q$'s derivatives.

To provide conservation of specific entropy when crossing transition
boundaries and proper behavior during the equilibrium mixed phase evolution
one needs to fulfill the following conditions (see Appendix, eq.(28)):
\begin{equation}
\frac {w_{Q}(T_d, \mu/3)}{w_H(T_d,\mu)} =
\frac {s_{Q}(T_d, \mu/3)}{s_H(T_d, \mu)} = \frac {n_{Q}(T_d,
\mu/3)}{n_H(T_d,\mu)}.
\end{equation}
\begin{equation}
\frac {w_{QGP}(T_{ch}, \mu/3)}{w_Q(T_{ch},\mu/3)} =
\frac {s_{QGP}(T_{ch}, \mu/3)}{s_Q(T_{ch}, \mu/3)} = \frac {n_{QGP}(T_{ch},
\mu/3)}{n_Q(T_{ch},\mu/3)},
\end{equation}
valid at $Q$-phase boundaries $T_d(\mu)$ and $T_{ch}(\mu)$.
Combining together (10)-(13) we get
certain {\it constraints} on $B_Q(\mu,T)$ function
instead of differential equation similar to (7). Namely:
\begin{itemize}
\item at the deconfinement boundary, $T(\mu)$ = $T_d(\mu)$:
\begin{equation}
(\frac {\partial B_Q}{\partial \mu})_{T=T_d(\mu)} =
( n^0_Q - n_H \frac {w^0_Q}{w_H})_{T=T_d(\mu)};
\end{equation}
\begin{equation}
(\frac {\partial B_Q}{\partial T})_{T=T_d(\mu)} =
( s^0_Q - s_H \frac {w^0_Q}{w_H})_{T=T_d(\mu)};
\end{equation}
\item at the chiral restoration boundary, $T(\mu)$ = $T_{ch}(\mu)$
\begin{equation}
(\frac {\partial B_Q}{\partial \mu})_{T=T_{ch}(\mu)}
= ( n^0_Q - n_{QGP} \frac {w^0_{Q}}{w_{QGP}})_{T=T_{ch}(\mu)};
\end{equation}
\begin{equation}
(\frac {\partial B_Q}{\partial T})_{T=T_{ch}(\mu)}
= ( s^0_Q - s_{QGP} \frac {w^0_{Q}}{w_{QGP}}))_{T=T_{ch}(\mu)};
\end{equation}
\item at $\mu =0$:
\begin{equation}
\quad B_Q(0,T) = B^0_Q = const
\end{equation}
\item inside the $Q$ phase region there is no special constraints on $B_Q$
(besides the Gibbs relation (8) which is valid for any $T$ and $\mu$),
and solution of the problem is not unique, thus
the function $B_Q(\mu, T)$ can be chosen {\it rather arbitrarily}.
\end{itemize}

In classical physics similar problems arise, eg., when simulating
soap films stretched on some hard contour \cite{plateau}
(so called two-dimensional Plateau problem).
Variation methods for such problems are well elaborated \cite{surface}.
We have used a numerical procedure providing
function $B_Q(\mu,T)$ which belongs to the
so called {\it minimal surface} class, securing local
minimum of the surface functional:
\begin{equation}
\int \int_{<Q>} d\mu dT
\sqrt{1+ \partial_\mu B_Q(\mu,T)^2 +
\partial_\mu B_Q(\mu,T)^2}
\end{equation}
with given boundary conditions (14)-(18).  Fulfillment of the
Gibbs relation (8) has been tested.

Note that this procedure fails near $T=0$ region where the accuracy of calculation
becomes worse; this case needs special investigation. However we are interested
mainly in high-$T$ region because this very case can be related to
modern experiments on heavy ion ultrarelativistic collisions where formation
of deconfined phase(s) seems to be rather probable.
\subsection{Results of numerical calculations}
\hspace*{.8cm} The procedure described has been fulfilled within HCM and MFA models for
the following parameter values:
\begin{center}
$B_{QGP}$ = 0.5 GeV/fm$^3$, \qquad $B^0_Q$ = 70 MeV/fm$^3$.
\end{center}
\par Fig.3a represents the crossection of the obtained $B_Q(\mu ,T)$ surface for
HCM by the deconfinement transition boundary $T=T_d(\mu)$ (solid line) and
the chiral transition boundary $T=T_{ch}(\mu)$ (dashed line).
The same curves for MFA model are presented in Fig.3b.
It is seen that $B_Q(\mu,T)$ remains practically unchanged for small
$\mu$ values $\mu \le 0.6 GeV$. For larger $\mu$,
$B_Q(\mu, T)$ value varies
within rather broad interval (relatively to its average value):
50 MeV/fm$^3$ $\leq$ $B_Q(\mu,T)$ $\leq$ 90 MeV/fm$^3$.
However the interval of $B_Q$ variation
is small as compared to difference between $B_{QGP}$ and $B^0_Q$.
Moreover, phase diagrams before and after $B_Q$ modification (see Fig.4 ) {\it do not
differ significantly}. The region of $Q$ phase existence remains rather broad
with varying $B_Q$ as well.

The modification procedure described above influences mainly the baryon
number density in the $Q$ phase (see Fig.5), corresponding entropy
density  corrections satisfy Gibbs relation and are relatively small
(everywhere except low $T$ region).

It deserves stressing that $B_Q(\mu ,T)$ differs for HCM and MFA models
reflecting intrinsic incorrectness of the models themselves:
$Q$ phase does play its role of intermediate state compensating
entire defects of $H$ phase EOS.
The boundary of chiral transition does not depend practically on peculiarities
of nucleon interaction in $H$ phase; this means that EOS of $QGP$ does not
remember the $H$-phase interactions. This seems to be quite reasonable.

\section {Summary and discussion}
\hspace*{.9cm}It has been shown that within DPTM the modification
$B_Q \rightarrow B_Q(\mu ,T)$ enables to provide proper behaviour of
thermodynamic functions for reversible equilibrium
phase transition, in particular,  the mixed phase scenario.
Phase diagram of three-phases matter is not practically changed.
The correction concerns mainly baryon number density inside the
intermediate $Q$-phase (its changes are not essential). The main
result of DPTM - existence of a broad corridor of $Q$ phase in $\mu-T$
plane - remains entirely valid.

Note that within DPTM modifications are necessary and considerable
only for sufficiently large chemical potential,
 $\mu \ge $ 0.6$\div$0.8 GeV.
For small $\mu $ (most interesting for experimental data analysis)  the
equilibrium character of deconfinement phase transition is almost
automatically saved, and the change of the chiral transition boundary
is negligible.

Let us stress that EOS of $QGP$ remains unchanged within the method used,
thus the problem of the direct transition $QGP \leftrightarrow H$
irreversibility remains as well.
But within DPTM the transition is to proceed through the intermediate $Q$
phase so that direct transition $H\leftrightarrow QGP$ should not occur normally.
However there still remains the possibility for $QGP$ to overcool (too fast)
below the critical temperature $T_{ch}$, then the {\it nonequilibrium }
and thus irreversible phase transition should occur.

It deserves mentioning that the interest to the problem of
specific entropy discontinuity
has been inspired mainly by recent experimental data
on heavy ion high energy collision \cite{exp} reporting a large value of
${\cal S/N_B}$ for hadrons resulting from the  collision.
In this very connection it has been pointed out \cite{raf} that
experimental data does agree with the value typical for $QGP$ and
could appear in experiment due to abrupt nonequilibrium phase transition
$QGP \rightarrow H$. However, it should be stressed that
the ratio {\cal $S/N_B$} calculated within DPTM for the resulting hadron gas
(see Fig.6) {\it is much higher than that for direct transition} (and almost
the same as in intermediate $Q$ phase).
It is connected with relatively low temperature of H-phase
formation, $T_d \simeq $ 140 MeV. Thus experimental data
could be as well described by {\it equilibrium} transition $Q \rightarrow H$
instead of abrupt direct transition.

In conclusion let us point out another problem concerning EOS uncertainties.
There were put out arguments \cite{mass} based on theoretical analysis of effective
Lagrangian that hadron (and valon as well) masses are
to decrease for temperature increasing (since the mass of any hadron
is believed to be proportional to the quark condensate to the order 1/3),
with the fastest decrease (down to zero)
being close to critical temperature of direct transition
in SPTM. Actually this decrease means that the deconfined phase with broken
chiral symmetry transforms {\it smoothly} into $QGP$ without the
(at least, first order) phase transition.
In accordance with such approach  DPTM is to be modified in such a way that
change of valon masses and corresponding change of bag pressure parameters
(which have to be connected with quark condensate and gluon condensate) are
taken into account. It seems natural that in this case
there occurs the only phase transition, the
deconfinement one, $H \leftrightarrow Q$ , while transformation of the
$Q$ phase into $QGP$ proceeds smoothly, with constituent quark masse
approaching zero. Then the problem of the deconfinement transition
irreversibility remains actual and needs special analysis.

\subsection*{Acknowledgement}
\hspace*{.8cm} The author is very much indebted to Professor E.L. Feinberg for helpful
discussions and encouraging this work. We are also very much obliged to
Professor H. Satz for helpful critical discussions and valuable remarks.
 The work is partially supported by RFFI grant N 96-02-19572.\\
\subsection*{Appendix }
\begin{small}
\hspace*{.9cm} Hydrodynamical equations describing the system evolution
are to be written for energy-
momentum tensor; they express the local conservation laws. Neglecting
dissipative effects, they are:
\begin{equation}
\partial_\mu T^{\mu\nu}(x)=0,\qquad
 T^{\mu\nu}(x)=[\epsilon(x)+p(x)]u^\mu(x) u^\nu(x)+p(x)g^{\mu\nu},
\nonumber
\label{a1}
\end{equation}
where $\epsilon$ is the energy density, $p$ the pressure
in the proper system of fluid element,
$g^{\mu\nu}$ the metric tensor (with $g^{00}=-1$)
and $u^\mu$ is the four velocity of the local flow,
$x$ being the coordinate of the fluid element in the 4-dimension space.
Conservation of the entropy and baryon number requires for:
\begin{equation}
\partial_\nu (u^\nu s) = 0,
\end{equation}
\begin{equation}
\partial_\nu (u^\nu n) = 0,
\end{equation}
where $s$ and $n$ being respectively the entropy density and baryon number
density in the proper system of fluid element.
In the case of one-dimensional scaling solution \cite{bjor} commonly used for qualitative
description of evolution of the matter created in heavy ion collisions,
local conservation of energy-momentum tensor takes the form \cite{rev}
 ($\tau$ being the proper time of fluid element):
\begin{equation}
 \frac {d \epsilon}{\epsilon + p} = - \frac {d \tau }{\tau}
\end{equation}
Until $p$ (as well as $T$) is {\it constant} during the mixed phase
the last equation presents
{\it enthalpy $w=\epsilon +p$ conservation law}. Local conservation of entropy and
baryon number is expressed in a similar form:
\begin{equation}
\frac {ds}{s} =- \frac{d\tau}{\tau},
\qquad \frac {dn}{n} = - \frac {d\tau }{\tau}
\end{equation}

Thus to provide a possibility for an equilibrium  transition process
through the mixed phase state the following conditions are required:
\begin{equation}
\frac {dw}{w} =  \frac {ds}{s}  =  \frac {dn}{n}  =  -\frac {d\tau}{\tau}
\end{equation}
\label{3=}

In the case of direct transition
the mixed phase state is described as a mixture of $H$  and $QGP$ fractions
taken at constant temperature $T_c$ and given chemical potential $\mu $:
\begin{equation}
w=w_H(T_c,\mu) \lambda(\tau) +  w_{QGP}(T_c, \mu/3)(1-\lambda(\tau)),
\end{equation}
$\lambda (\tau)$ being the share of
the $H$-phase admixture which changes during the mixed phase expansion from
zero (for pure $QGP$ phase) to the unity (for pure hadronic matter). The
same representation is taken for all additive thermodynamic variables ($s,
n, \epsilon $).
Thus, (25) and (26) combining together lead to the following requirement:
\begin{equation}
\frac {w_{QGP}(T_c, \mu/3)}{w_H(T_c,\mu)} =
\frac {s_{QGP}(T_c, \mu/3)}{s_H(T_c, \mu)} = \frac {n_{QGP}(T_c,
\mu/3)}{n_H(T_c,\mu)}.
\end{equation}

In general case of any first order equilibrium phase transition
at some transition temperature $T_{tr}$
corresponding requirement takes the form:
\begin{equation}
\frac {w^+(T_{tr}, \mu)}{w^-(T_{tr},\mu)} =
\frac {s^+(T_{tr}, \mu)}{s^-(T_{tr}, \mu)} = \frac {n^+(T_{tr},
\mu)}{n^-(T_{tr},\mu)},
\end{equation}
where (+) and (-) refer to thermodynamic variables above and below the
transition boundary $T_{tr}(\mu)$.

Let us stress that for equilibrium mixed phase scenario it is necessary
and sufficient to fulfill the conditions (28)
{\it only at transition boundaries}, but not for any $T$ and $\mu$.
Besides, for $\mu =0$ (when $n$=0) and $T$=0 (when $s$=0) cases
these conditions are satisfied automatically due to common thermodynamic
Gibbs relation (8).

It deserves stressing (and can be easily shown) that fulfillment of the
conditions (28) provides automatically securing of the Gibbs relation (8)
 in one of the neighboring phases if it is valid for the another one.
\end{small}
\newpage
\subsection*{Figure Captions}
\noindent {\bf Figure 1.} Phase diagram in $\mu-T$ plane for
nucleon interaction description within
HCM (solid) and MFA (solid with symbols).
Dashed lines correspond to the direct transition in SPTM.\\

\noindent {\bf Figure 2a.} Ratio of specific entropy values above (+) and
below (-) the {\it direct} transition,
 $({\cal  S/N_B})^+:({\cal S/N_B})^-$,
as a function of baryonic chemical potential $\mu$ along the transition
boundary.
Calculated for HCM (solid) and MFA (solid-symbols).\\
{\bf Figure 2b.} The same ratio as in Fig. 2a at the deconfinement (solid)
and chiral (dashed) transition boundaries; curves for MFA model are indicate
by symbols. Note the different ordinate scales.\\

\noindent {\bf Figure 3a.} Crossection of the obtained $B_Q(\mu ,T)$ surface
by the deconfinement transition boundary $T=T_d(\mu)$ (solid line) and the
chiral transition boundary $T=T_{ch}(\mu)$ (dashed line)
as a function of baryonic chemical potential $\mu$.
Calculated within HCM. \\
{\bf Figure 3b.} The same as in Fig. 3a calculated within MFA.\\

\noindent {\bf Figure 4a.} Phase diagram within HCM for
$B_Q$=70 MeV/fm$^3$ =Const (dashed) and modified $B_Q(\mu, T)$ (solid).
Short-dashed lines correspond to the limiting values of modified $B_Q(\mu,T)$:
 $B_Q$=50 MeV/fm$^3$ and $B_Q$=90 MeV/fm$^3$ for deconfinement and chiral
 transitions respectively.\\
{\bf Figure 4b.} The same as in Fig. 4a for MFA. \\

\noindent {\bf Figure 5a.} Baryon number density in $Q$ phase $n_Q$ as a function of
baryonic chemical potential $\mu$ at deconfinement (d) and chiral (ch)
transition boundaries calculated for $B_Q$=Const (dashed) and modified
$B_Q(\mu, T)$ (solid) for HCM.\\
{\bf Figure 5b.} The same as in Fig. 5a for MFA.\\

\noindent {\bf Figure 6.} Entropy per baryon in $H$ phase near the deconfinement
in DPTM (solid) and direct in SPTM (dashed) transition boundaries as function
of quark fugacity $\lambda = exp (\mu_q/T)$.  Short-dashed line corresponds
to the same ratio in $QGP$ phase.  Calculated within HCM.\\

\newpage 
~\\
\vspace*{5.5cm}
   \begin{figure}
   \vspace*{-.0cm}
   \hspace*{.0cm}
   \epsfxsize=6.truein
   \epsfysize=6.4truein
   \epsffile{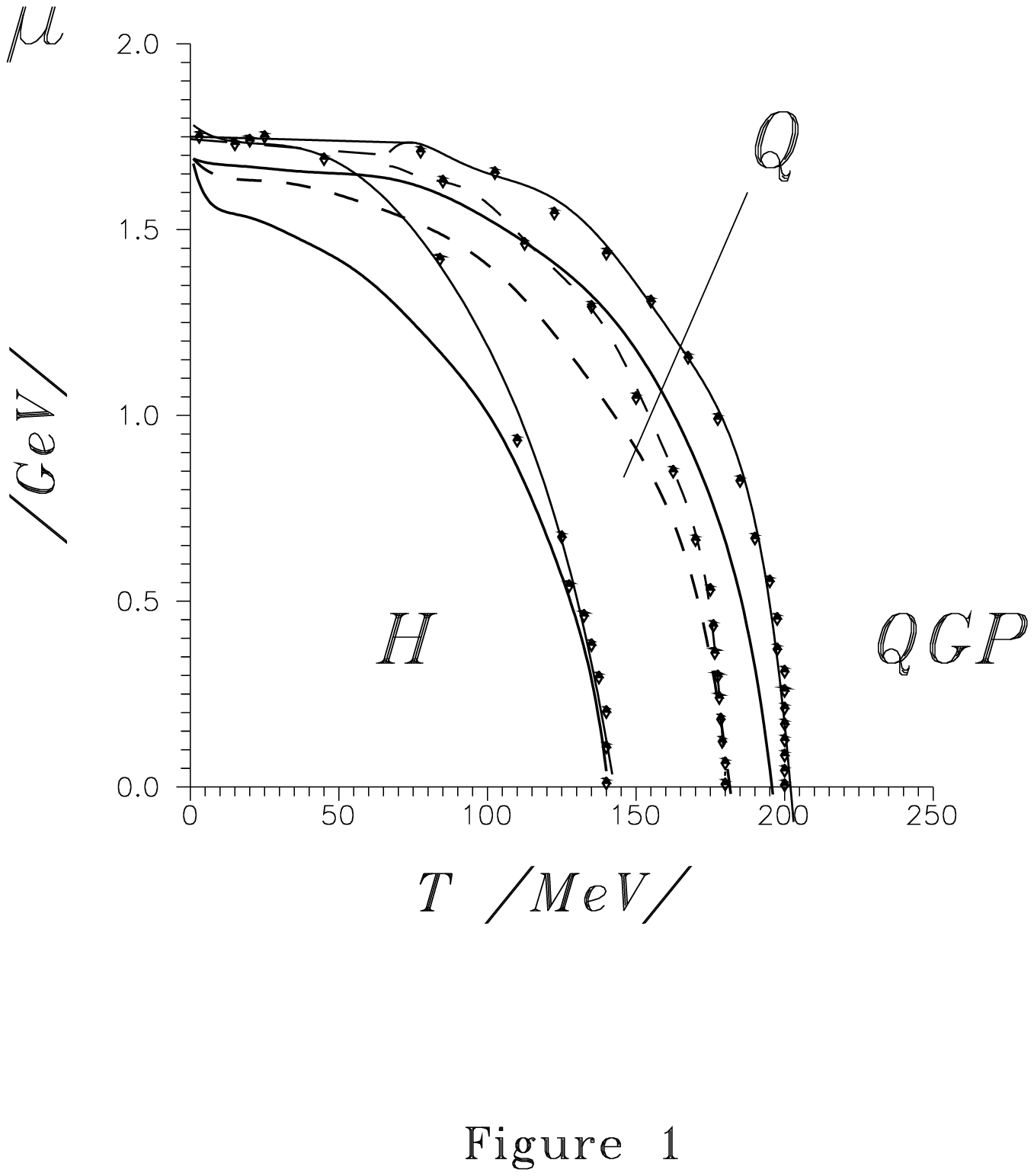}
\end{figure}
\newpage
\begin{figure}
   \vspace*{1.5cm}
   \hskip 2.7truein
   \epsfxsize=5.4truein
   \epsfysize=6.truein
   \epsffile{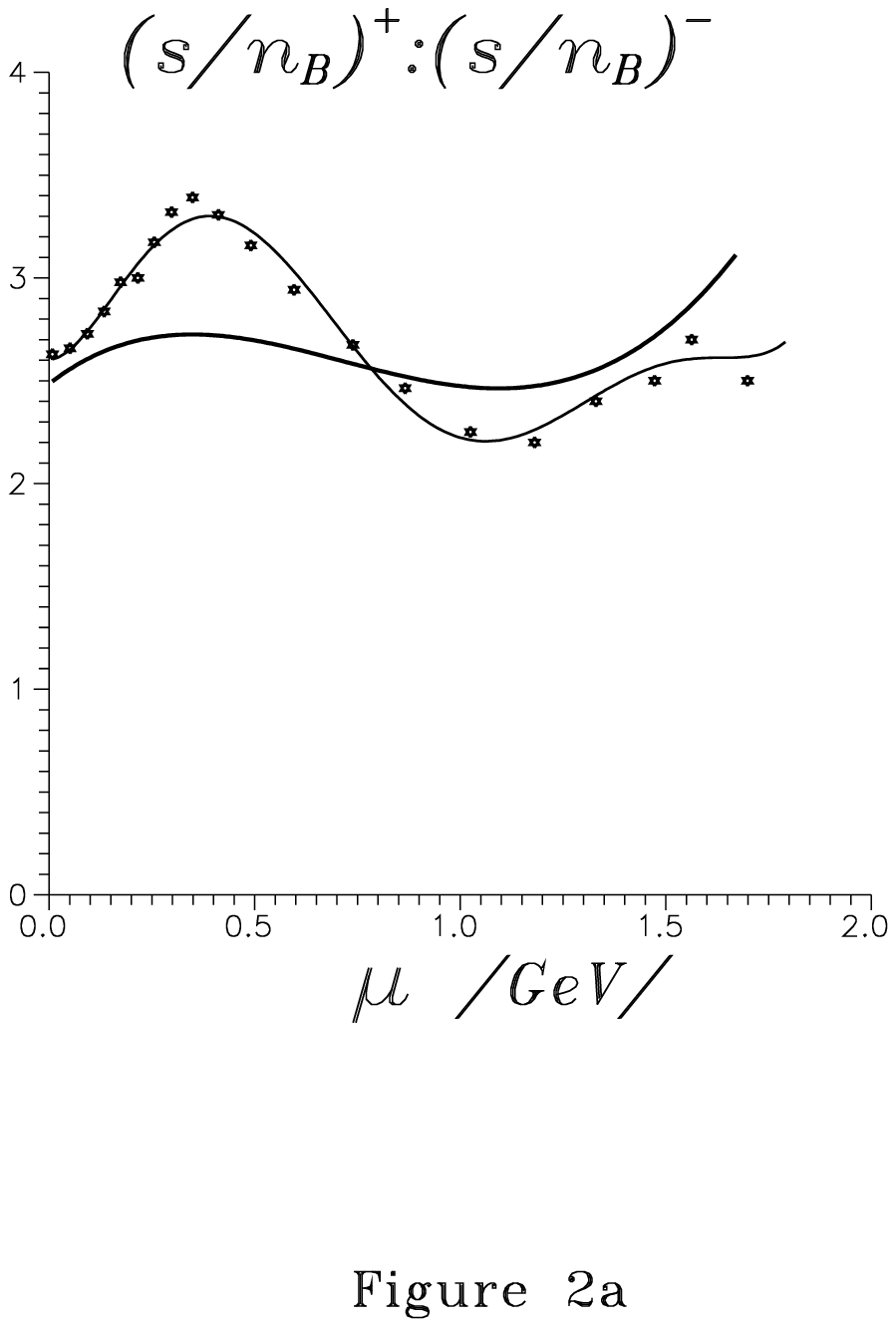}
   \end{figure}

   \vspace*{-4.5cm}

   \hskip 2.7truein
   \hspace*{-8.cm}
   \epsfxsize=5.4truein
   \epsfysize=6.truein
   \epsffile{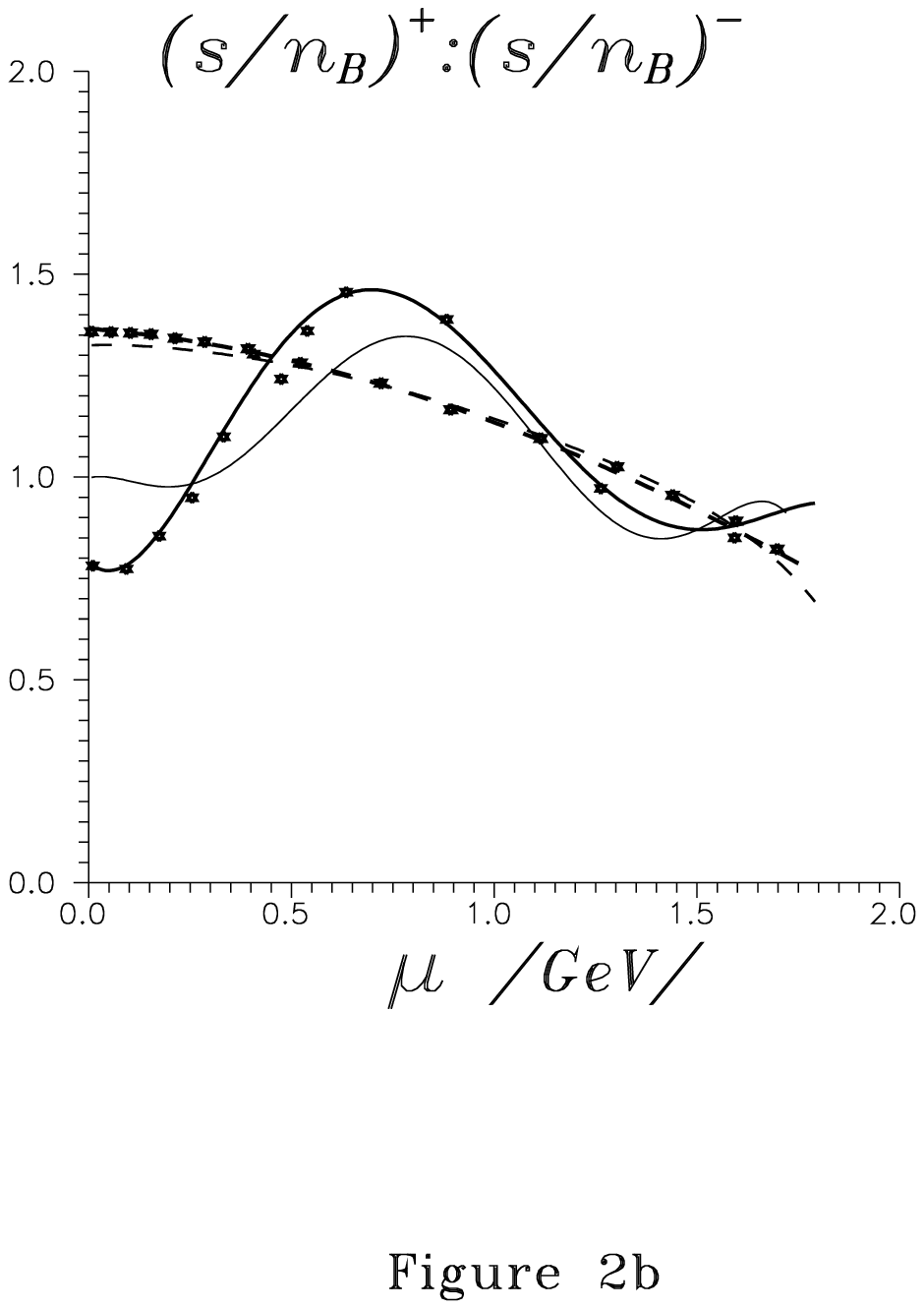}

\newpage
\begin{figure}
   \vspace*{2.cm}
   \hskip 2.7truein
   \epsfxsize=5.4truein
   \epsfysize=5.4truein
   \epsffile{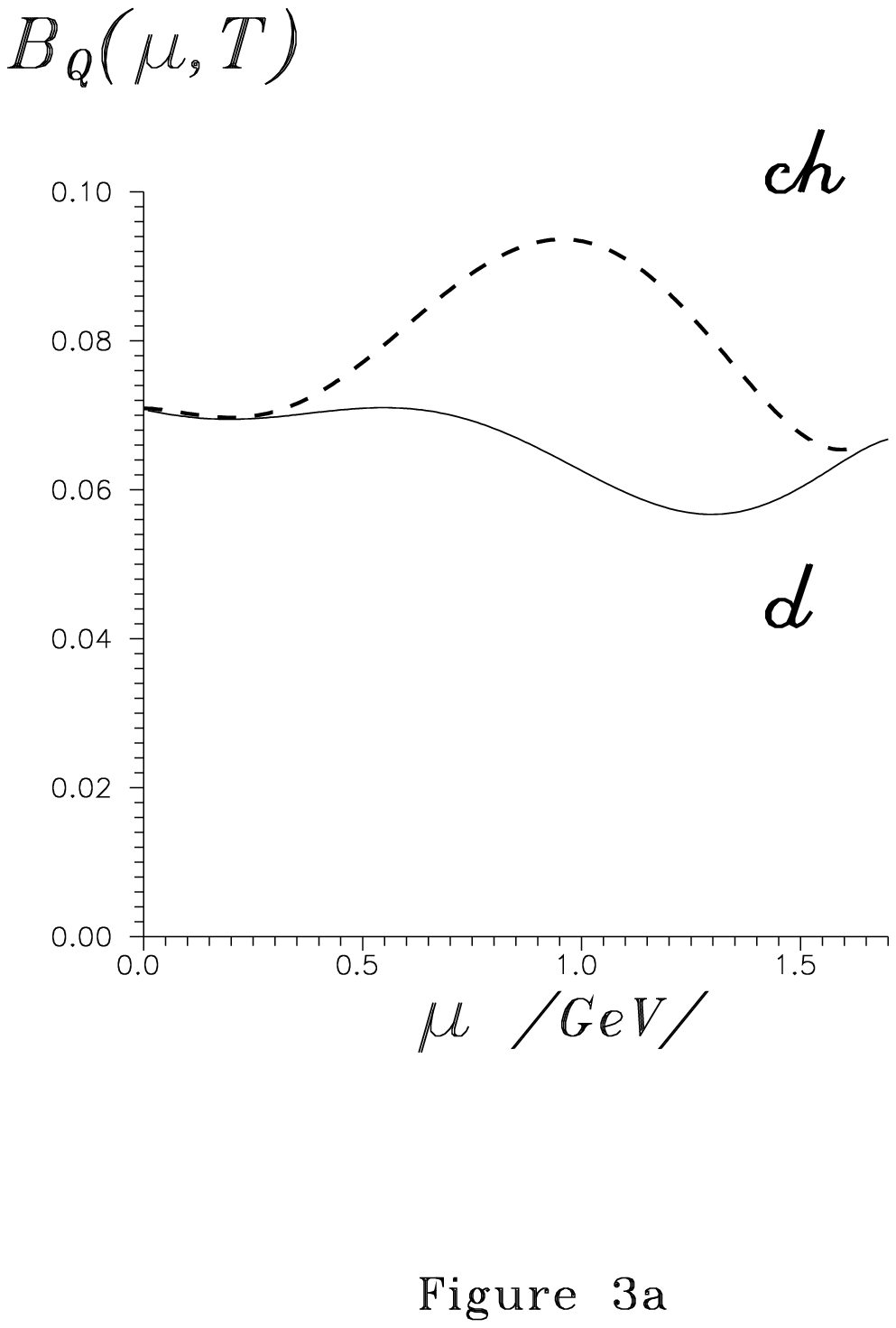}
   \end{figure}

   \vspace*{-4.5cm}
   \hskip 2.7truein
   \hspace*{-8.cm}
   \epsfxsize=5.4truein
   \epsfysize=5.4truein
   \epsffile{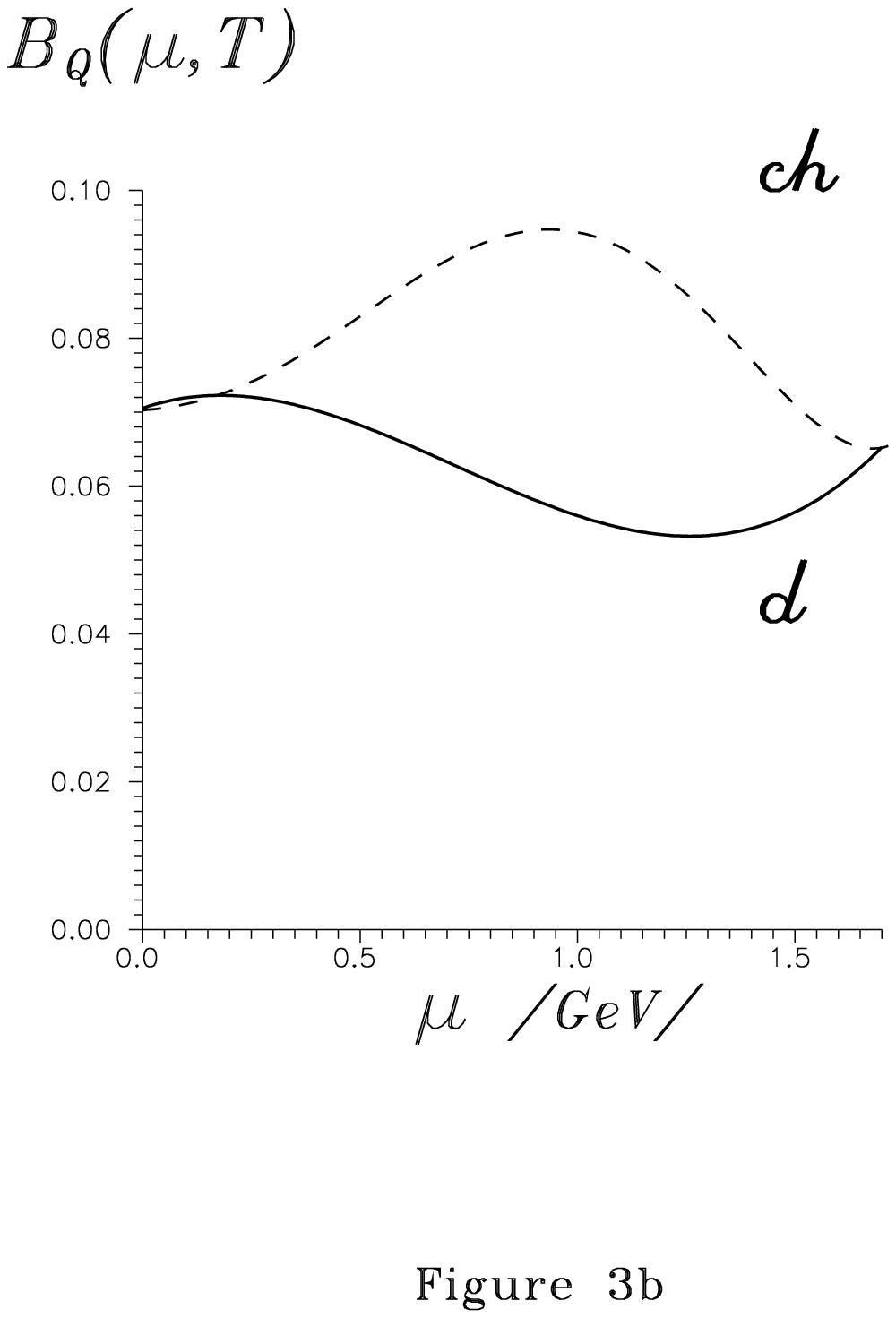}
\newpage
\begin{figure}
   \vspace*{1.5cm}
   \hskip 2.7truein
   \epsfxsize=5.4truein
   \epsfysize=6.truein
   \epsffile{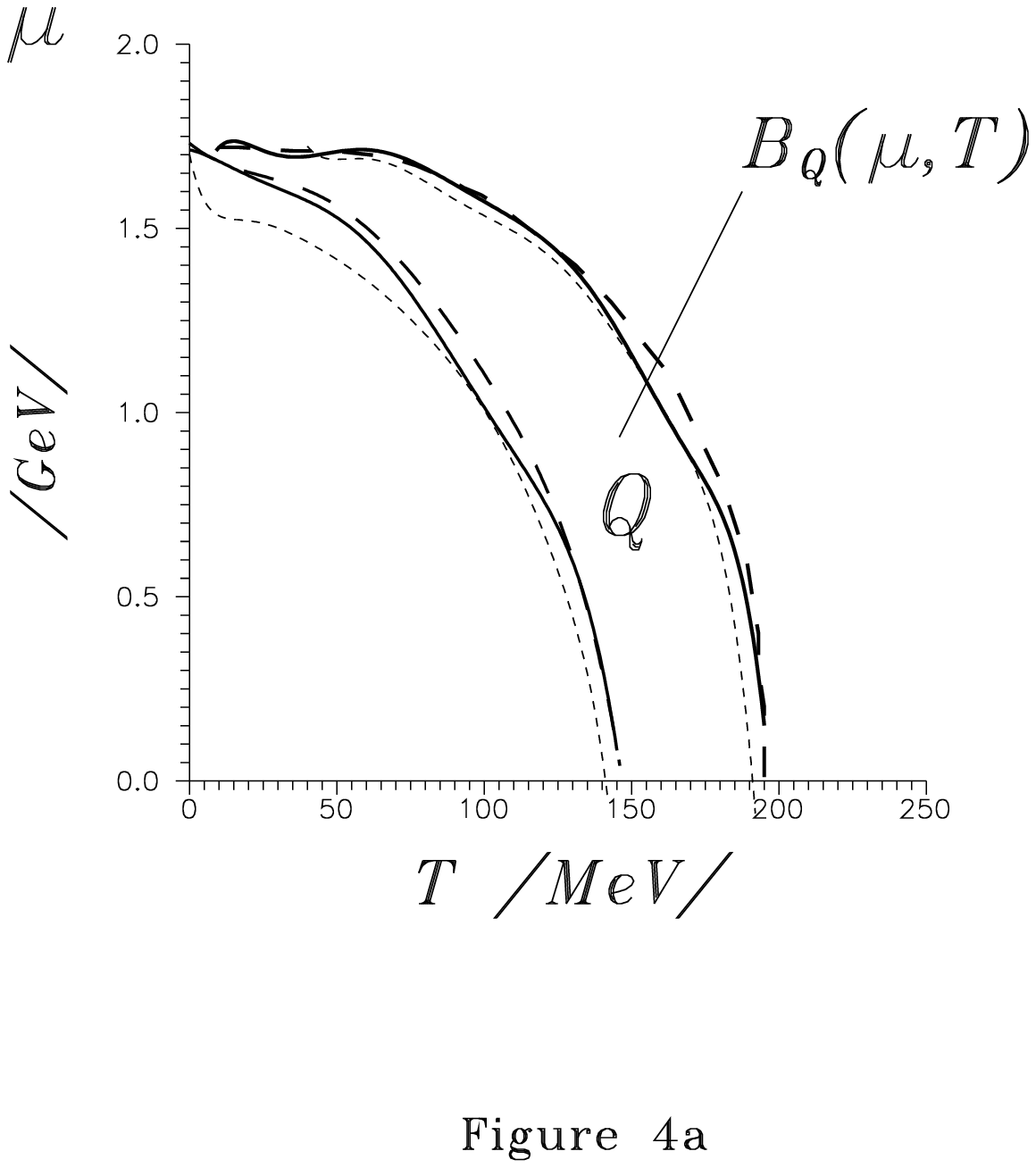}
   \end{figure}

   \vspace*{-4.5cm}
   \hskip 2.7truein
   \hspace*{-8.cm}
   \epsfxsize=5.4truein
   \epsfysize=6.truein
   \epsffile{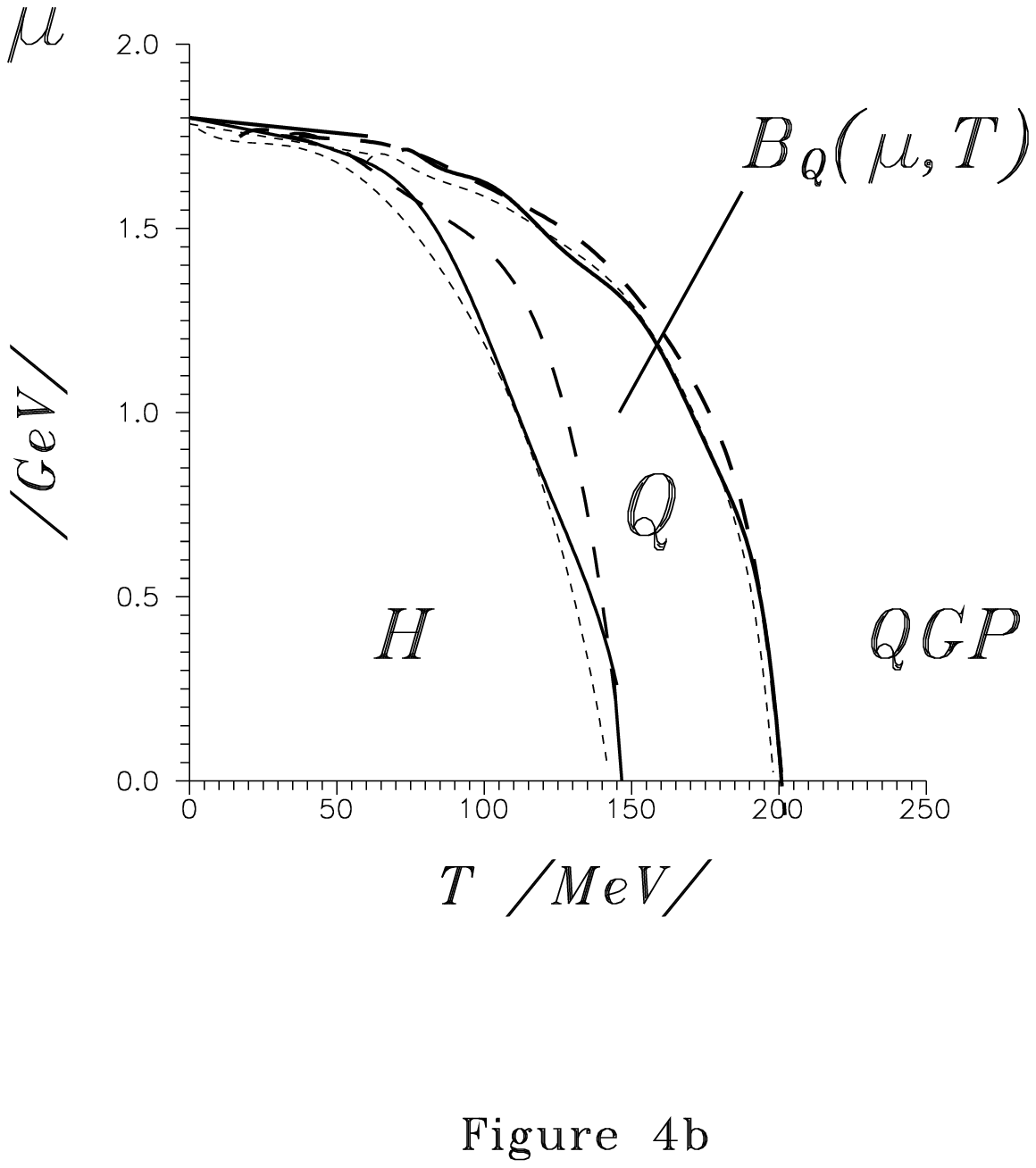}

\newpage
\begin{figure}
   \vspace{1.5cm}
   \hskip 2.7truein
   \epsfxsize=5.4truein
   \epsfysize=6.truein
   \epsffile{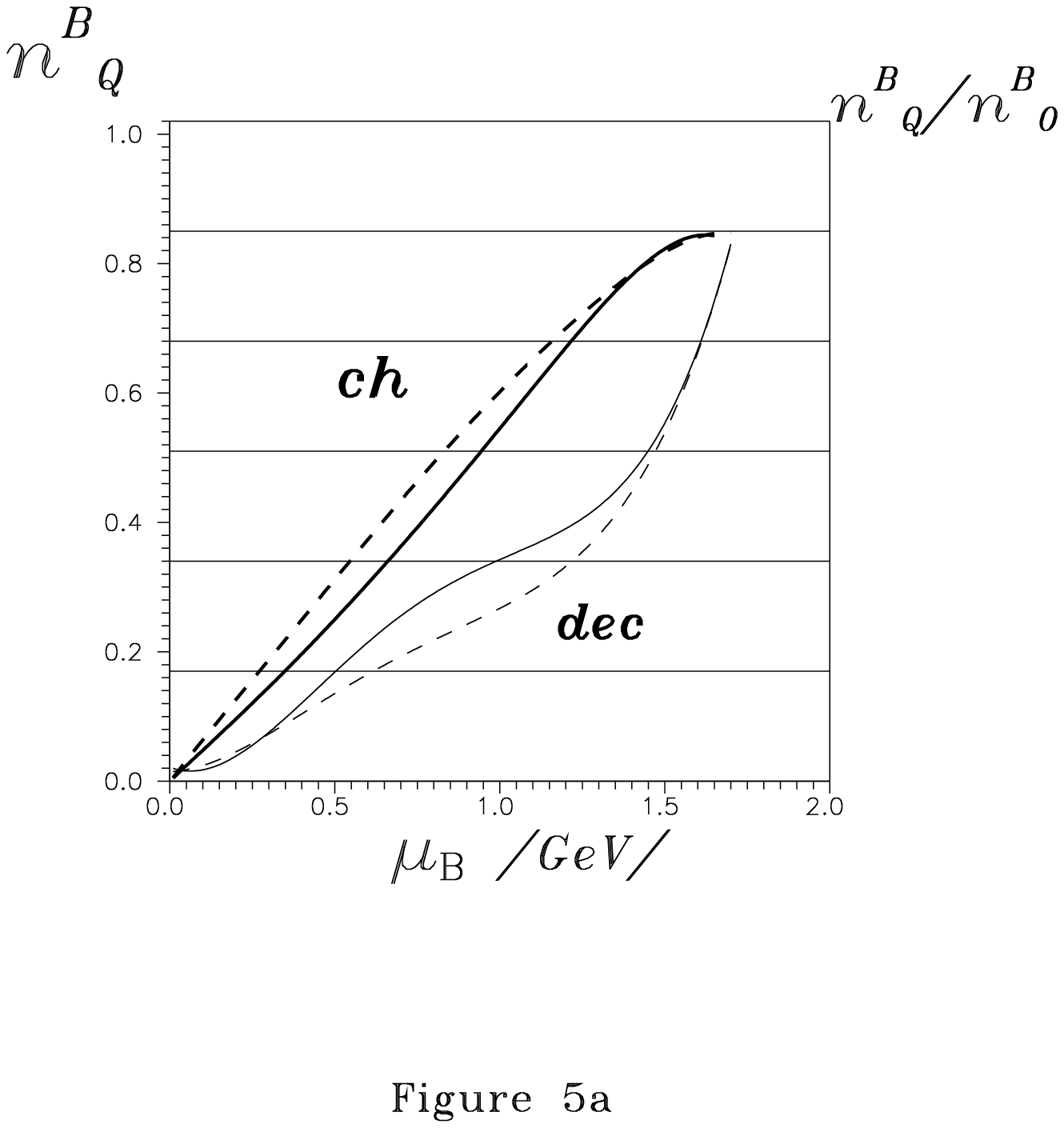}
   \end{figure}

   \vspace*{-4.5cm}
   \hskip 2.7truein
   \hspace*{-8.cm}
   \epsfxsize=5.4truein
   \epsfysize=6.truein
   \epsffile{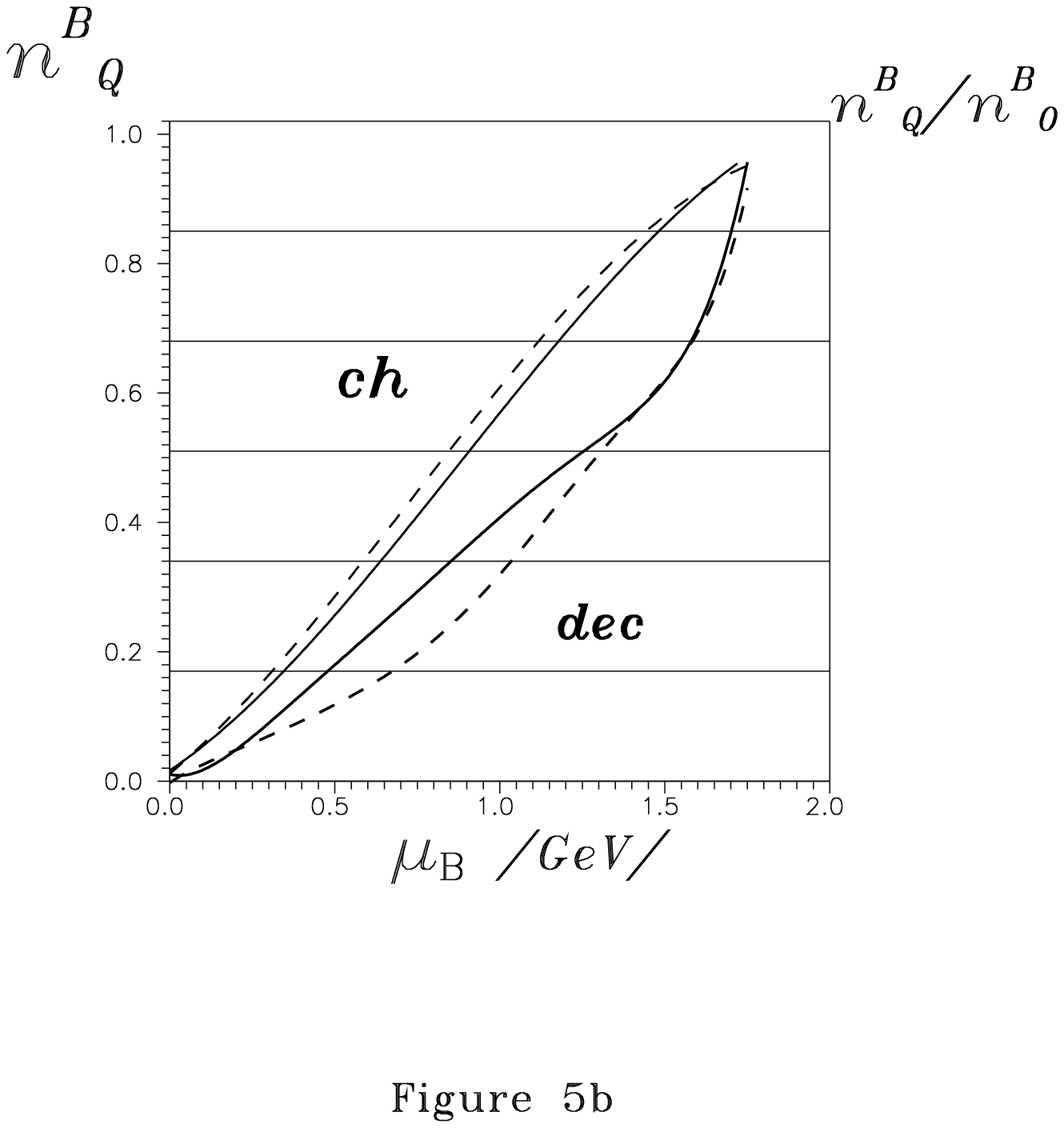}
\newpage 
   \begin{figure}
   \vspace*{-3.cm}
   \epsfxsize=6.1truein
   \epsfysize=7.truein
   \epsffile{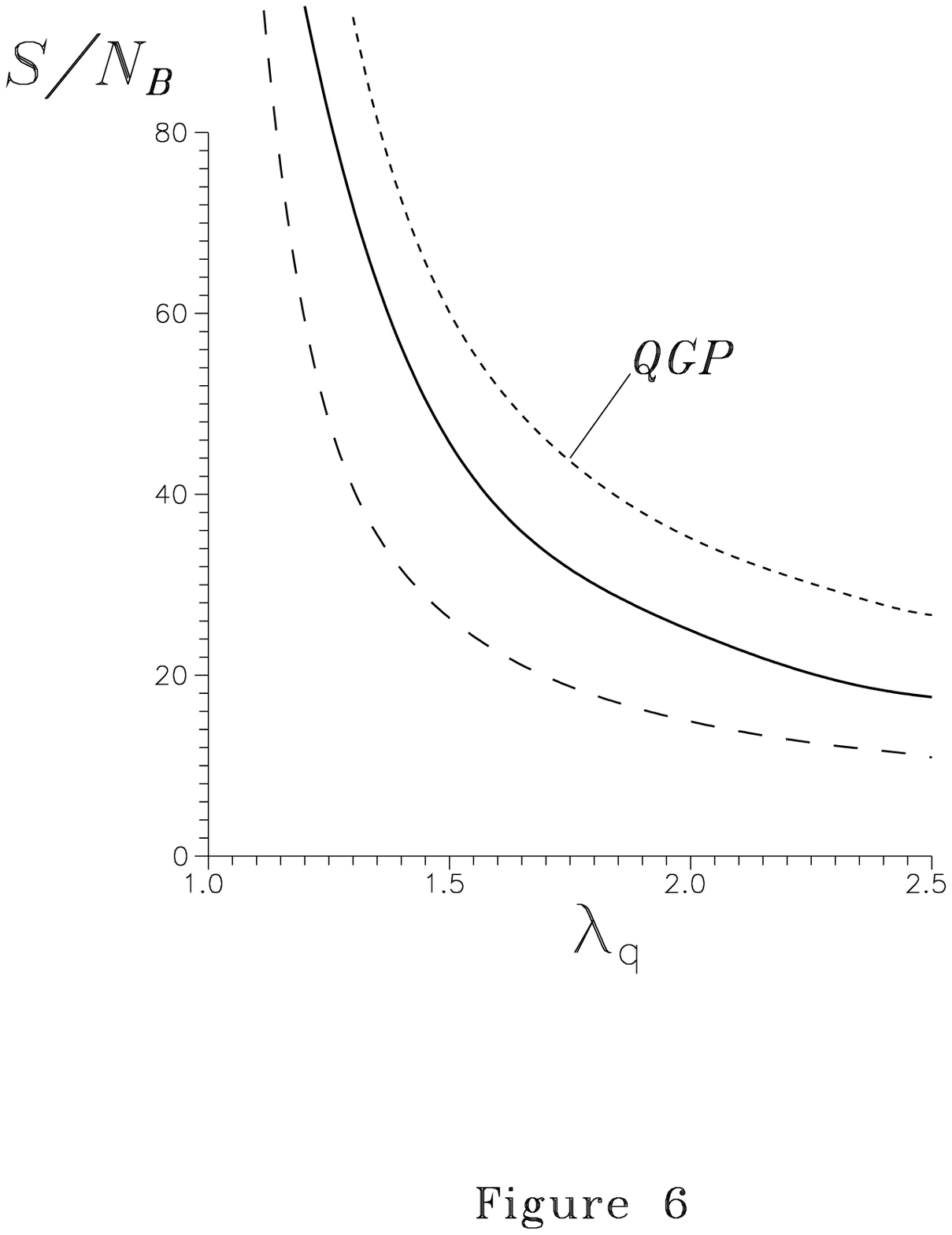}
\end{figure}


\end{document}